\documentclass[twocolumn]{aa} 
\usepackage{graphicx} 

\begin{document}

\newcommand{\hi}{H{\small I}}
\newcommand{\mgi}{Mg{\small I}}
\newcommand{\oiii}{[O{\small III}]}
\newcommand{\hb}{H$\beta$}
\newcommand{\kms}{$\rm{km\,s^{-1}}$} 
\newcommand{\mjb}{mJy beam$^{-1}$}

\def\ma2{$\rm{mag\,arcsec^{-2}}$} 
\def\kmsMpc{$\rm{km\,s^{-1}}$\,Mpc}
\def\msun{M$_{\sun}$} 
\def\lsun{L$_{\sun}$}
\def\mjyb{mJy~beam$^{-1}$}

\title{NGC\,5719/13: interacting spirals forming a counter--rotating stellar disc
\thanks{Based on observations carried out at the European Southern
 Observatory (ESO 69.B-0515)}}
\author{
	D. Vergani 	\inst{1,2}
\and	A.~Pizzella	\inst{3}
\and    E.M.~Corsini    \inst{3,4}
\and    W.~van Driel    \inst{2}
\and    L.M.~Buson      \inst{5}
\and   R.-J.~Dettmar   \inst{6}
\and   F.~Bertola      \inst{3}
 }

 \offprints{D. Vergani}
 \mail{daniela@lambrate.inaf.it}

 \institute{
      INAF -- IASF Milano, via Bassini 15, I-20133 Milano, Italy
\and  
Observatoire de Paris, Section de Meudon, GEPI, 5 place Jules Janssen, F-92195 Meudon Cedex, France
\and  
Dipartimento di Astronomia, Universit\`a di Padova, vicolo dell'Osservatorio 3, I-35122 Padova, Italy
\and  
Scuola Galileiana di Studi Superiori, via VIII Febbraio 2, I-35122 Padova, Italy
\and
INAF -- Osservatorio di Padova, Vicolo dell'Osservatorio 5, I-35122 Padova Italy
\and  
Astronomisches Institut der Ruhr-Universit\"at Bochum, D-44780 Bochum, Germany
}

\date{Received 19 September 2006 / Accepted 9 November 2006}

% \abstract{}{}{}{}{} 

\abstract
% context heading (optional), leave it empty if necessary 
{When a galaxy acquires material from the outside, it is likely that the
resulting angular momentum of the accreted material is decoupled from that of 
the pre-existing galaxy. The presence of stars counter-rotating with respect 
to other stars and/or gas represents an extreme case of decoupling.}
{NGC\,5719, an almost edge-on Sab galaxy with a prominent skewed dust lane,
  shows a spectacular on-going interaction with its face-on Sbc companion
  NGC\,5713. Observations of such interacting systems provide insight into the
  processes at work in assembling and reshaping galaxies.} 
{Studies were made of the distribution and kinematics of neutral hydrogen in
  the NGC\,5719/13 galaxy pair and the ionised gas and stellar kinematics
  along the major axis of NGC\,5719.} 
{Two \hi\ tidal bridges that loop around NGC\,5719 and connect to 
NGC\,5713, and two \hi\ tidal tails departing westward from 
NGC\,5713 were detected. There is a correspondence between the \hi\ condensations 
and the location of clumps of young stars within and outside the disc of NGC\,5719.
The low-mass satellite PGC\,135857 at the tip of the northern tail was
detected in \hi, and is likely a by-product of the interaction.
The neutral and ionised hydrogen in the disc of NGC\,5719 are 
counter-rotating with respect to the main stellar disc. 
The counter-rotating stellar disc contains about 20\% of the stars 
in the system, and has the same radial extension as the main stellar disc. 
This is the first interacting system in which a counter-rotating stellar 
disc has been detected.}
{The data support a scenario where \hi\ from the large reservoir available in 
the galaxy's surroundings was accreted by NGC\,5719 onto a retrograde orbit 
and subsequently fuelled the in-situ formation of the counter-rotating 
stellar disc.}

\keywords{galaxies: individual (NGC\,5713, NGC\,5719) --- 
galaxies: interactions --- 
galaxies: kinematics and dynamics --- galaxies: spiral} 

\authorrunning{D. Vergani et al.} 
\titlerunning{NGC\,5719/13: interacting spirals} 
\maketitle

\section{Introduction}
\label{sec:introduction}
 
Strong interactions between galaxies often produce huge tidal tails
consisting of both cold gas and stripped, or even newborn stars, as
shown by numerical simulations as well as by detailed observations (see
Schweizer 2000 for a review). 
Gas plays a crucial role in such interactions, because of its
dissipative nature. When a galaxy acquires material from outside, it
is likely that the resulting angular momentum of the acquired material
is decoupled from that of the pre-existing galaxy. The
presence of stars counter-rotating with respect to other stars and/or
gas represents a truly extreme case of decoupling. 
This phenomenon has been detected in the central regions 
of a large number of galaxies, with morphological types ranging
from ellipticals to S0's and spirals (see Bertola \& Corsini 1999 
for a review). Although it is usually invoked as the signature of a past
external event, some attempts have been made to explain special cases of
stellar counter-rotating discs as due to a self-induced phenomenon in
non-axisymmetric potentials (Evans \& Collett 1994; Athanassoula et al. 1996;
Wozniak \& Pfenninger 1997).   

{\it Extended} stellar counter-rotation in discs of S0's and spirals appears 
to be a rare phenomenon (Kuijken et al. 1996; Kannappan \& Fabricant 2001; 
Falc{\'o}n-Barroso et al. 2006). It has been argued that the retrograde 
acquisition of small amounts of external gas can give rise to counter-rotating 
gaseous discs in gas-poor S0's only, while in
gas-rich spirals the newly acquired gas is swept away by the
pre-existing gas. Counter-rotating gaseous and stellar discs in
spirals are formed only from, respectively, the retrograde acquisition of 
amounts of gas larger than the pre-existing gas content, and subsequent 
star formation (Pizzella et al. 2004).

To date only few cases of extended counter-rotating stellar discs have been observed.
The S0 galaxy NGC\,4550 has two co-spatial counter-rotating stellar
discs, one of which is co-rotating with the gaseous one (Rubin et al. 1992).
The two discs have exponential surface brightness profiles with the
same central surface brightness and scale lengths (Rix et al. 1992), 
but different thickness (Emsellem et al. 2004).
In the early-type spirals NGC\,4138 (Jore et al. 1996) and NGC\,7217
(Merrifield \& Kuijken 1994) about $20$--$30\%$ of the disc stars
reside in retrograde orbits. These counter-rotating stellar discs have
the same spatial extent but a lower surface brightness than the main 
stellar disc. In NGC\,7217 the gaseous disc rotates in the
same direction as the main stellar disc. The contrary is true for
NGC\,4138, whose counter-rotating disc is currently forming stars.
The Sa galaxy NGC\,3593 (Bertola et al. 1996) is composed of a radially more
concentrated stellar disc that contains about $20\%$ of the total luminous
mass and dominates the stellar kinematics in the inner kpc, and a
counter-rotating stellar disc that is radially more extended and dominates the
outer kinematics. The two discs have exponential luminosity profiles with
different scale lengths and central surface brightnesses. The gaseous disc
co-rotates with the more concentrated disc, which is characterized by a high
star formation rate (Corsini et al. 1998; Garc{\'i}a-Burillo et al. 2000). 

The NGC\,5719/13 pair is the subject of the present paper.
NGC\,5719 is a highly inclined Sab galaxy system with a complex,
asymmetric morphology. It is characterized by a prominent dust lane 
skewed by about 25\degr\ with respect to the optical main body, and a 
chain of faint condensations in the north-western section of 
the outermost disc. Although it has been classified as a barred
galaxy by de Vaucouleurs et al. (1991, hereafter RC3), surface photometry 
shows no clear sign of a bar (Vergani et al. 2006).  
Its face-on Sbc companion NGC\,5713 has an overall undisturbed look in 
the optical, but it shows a conspicuous bar-like distortion pointing 
toward NGC\,5719 in the near-infrared (Mulchaey et al. 1997).

The pair belongs to the rich LGC\,386 group (Garcia 1993). 
The projected separation between the two galaxies is 77 kpc (11\farcm5) 
and their systemic velocities differ by about 160~\kms. Although their 
relative proximity and peculiar morphologies already suggested they are 
interacting (de Vaucouleurs \& de Vaucouleurs 1964), it is only in the 
\hi\ line images (Sect. \ref{sec:hi_morphology}) that we can really 
appreciate the amount and extent of the displaced and accreted material.
Some basic properties of the two galaxies are given in Table \ref{tab:basic}.

In this paper we report the detection of a counter-rotating stellar
disc in the spiral galaxy NGC\,5719 which is interacting with its
close companion galaxy NGC\,5713. The pair offers a unique opportunity 
to study the effects of an on-going acquisition event in the framework 
of the origin of counter-rotating stellar discs.
The paper is organized as follows. 
The optical and radio observations and data reduction are described in 
Sect.~\ref{sec:observations}. The results are presented in 
Sect.~\ref{sec:results} and discussed in Sect.~\ref{sec:discussion}, 
and our conclusions are given in Sect.~\ref{sec:conclusions}.

\begin{table} 
\begin{center}
\caption{Basic parameters of the NGC\,5719/13 galaxy pair.}
\label{tab:basic} 
\hspace*{-0.5cm} \begin{tabular}{lll}
\hline
\noalign{\smallskip}
         & NGC\,5719 & NGC\,5713 \\
\noalign{\smallskip}
\hline
\noalign{\smallskip}
$\alpha$ (J2000)    & 14$\rm{^h}$~40$\rm{^m}$~56\fs8 & 14$\rm{^h}$~40$\rm{^m}$~11\fs6\\
$\delta$ (J2000)    & $-$00\degr~19\arcmin~04\arcsec & $-$00\degr~17\arcmin~26       \\
Type                & SABab(s) pec                   & SABbc(rs) pec                 \\
$D_{25}$ (kpc)      & 21.8                           & 20.2                          \\ 
$i$      (\degr)    & 70                             & 27                            \\
P.A. (\degr)        & 96                             & 12                            \\
$V_{\it sys}$ (\kms)& $1741\pm8$                     & $1899\pm8$                    \\
$M^0_{B,T}$         & $-19.40$                       & $-20.43$                      \\
\hline
\noalign{\smallskip}  
\noalign{\smallskip}  
\noalign{\smallskip}  
\noalign{\smallskip}  
\end{tabular}
\end{center}
\begin{minipage}{8.5cm}
{\small \bf NOTES} --
The centre position, morphological type and optical diameter at a surface
brightness level of 25 $B-$mag arcsec$^{-2}$ are taken from the RC3.
The inclination is derived as $\cos^{2}{i}\,=\,(q^2-q_0^2)/(1-q_0^2)$. The
observed axial ratio $q$ is taken from the RC3 and an intrinsic flattening of
$q_0=0.11$ (Guthrie 1992) was assumed.
The position angle of optical major axis is taken from HyperLeda.
The heliocentric velocity derived from \hi\ data is from Vergani et al. (2006). 
The absolute total blue magnitude is corrected for inclination and extinction
and is taken from the RC3.
\end{minipage}
\end{table}

% ================ HI COLUMN DENSITY ===================
\begin{figure*}[t!]
\begin{center}
\leavevmode
\vskip -6cm
\hspace*{-1cm}\includegraphics[angle=0,width=21.5cm]{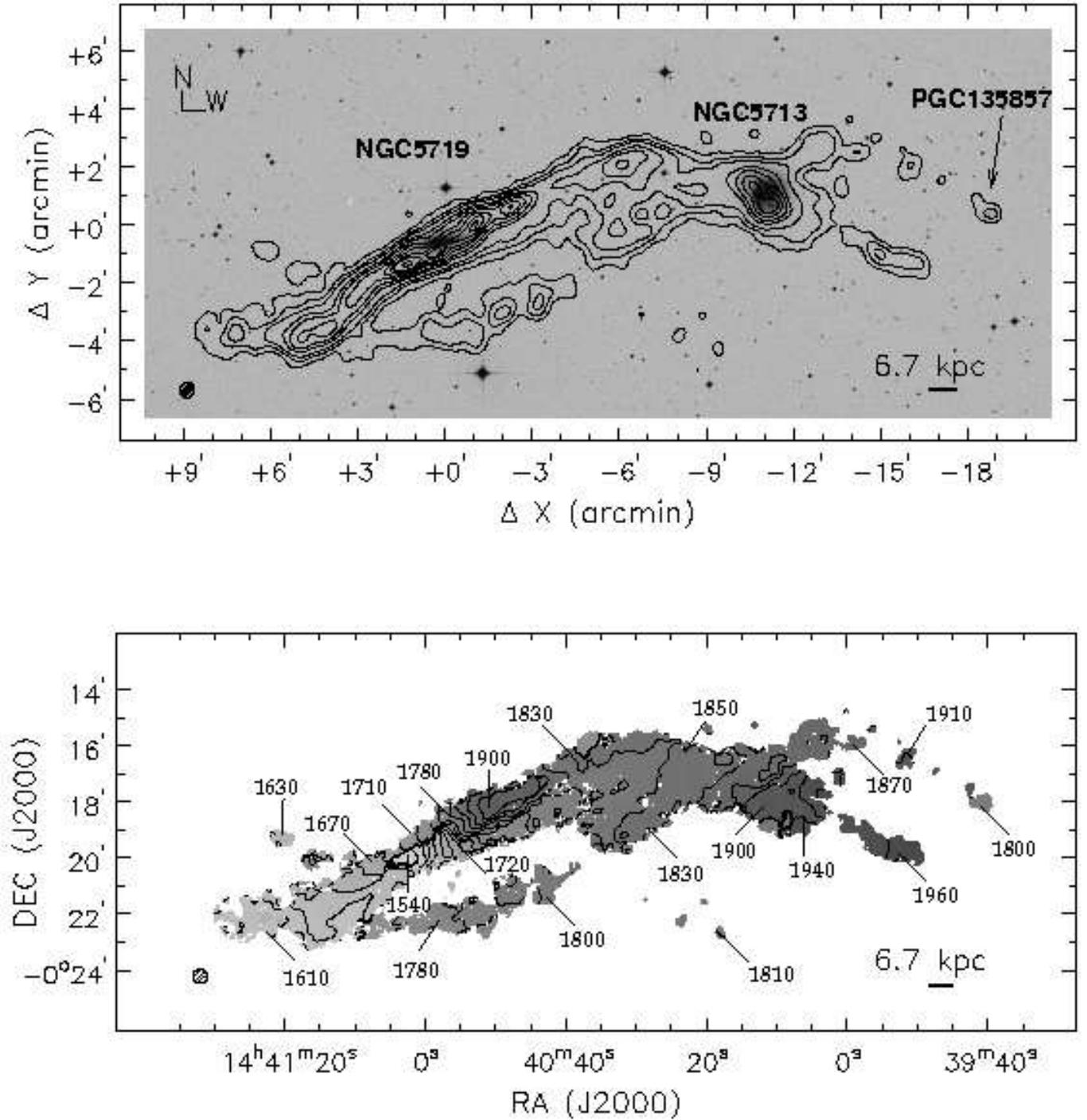}
\end{center}
\vskip -0.5cm
\caption{({\it Top}) Contour map of the \hi\ column density distribution of
  NGC\,5719/5713 superimposed on an optical (DSS2) image. 
The lowest contour level is at $7.0\times 10^{19}$ atoms cm$^{-2}$, and the
increment is $2.4\times10^{20}$ atoms cm$^{-2}$.
The spatial resolution is $30\arcsec\times30\arcsec$. 
1\arcmin\ is $\sim 6.7$~kpc at the assumed distance of 23.2\,Mpc. 
({\it Bottom}) \hi\ velocity field with contours and grey scales 
from 1500~\kms~ (approaching: eastern side, light shading) to 
2000~\kms~ (receding: western side, dark shading), in increments of 25~\kms.}
\label{fig:mom0velfi}
\end{figure*}

\section{Observations and data reduction}
\label{sec:observations}

\subsection{Radio observations}
 
We observed the NGC\,5719/13 pair with the Very Large Array 
(VLA) radio synthesis telescope in its C configuration in April 2000 
as part of a study of the kinematics of galaxies with box/peanut 
shaped bulges (Vergani et al. 2006). 
The galaxies were observed for 8~hours in the 4ABCD spectral line 
mode with two partially overlapping spectral bands, each with 64 channels of
24.4~kHz width and a total bandwidth of 2.685 MHz. The baselines range between
35~m and 3400~m.
The data cube used for further analysis has 109 velocity channels with a
velocity resolution of 5.15~\kms\ that cover a velocity range of $\sim
562$~\kms, an HPBW of $15\farcs5 \times 14\farcs1$, and an rms noise level
per channel of 0.53~\mjb.

A contour map of the \hi\ column density distribution smoothed to a resolution
of $30\arcsec \times 30\arcsec$ is shown superimposed on an optical image from
the Digitized Sky Survey in Fig.~\ref{fig:mom0velfi}. The velocity field
(Fig.~\ref{fig:mom0velfi}) was obtained by the modified envelope-tracing
method, which describes the kinematics in edge-on galaxies better than a
simple first moment analysis.
We refer to Vergani et al. (2006) for further details on the data reduction
and velocity field determination.

% ======== 3D VISUALIZATION ========
\begin{figure*}[t!]
\begin{center} 
\leavevmode
\vskip 0.5cm
\includegraphics[angle=0,width=18.0cm]{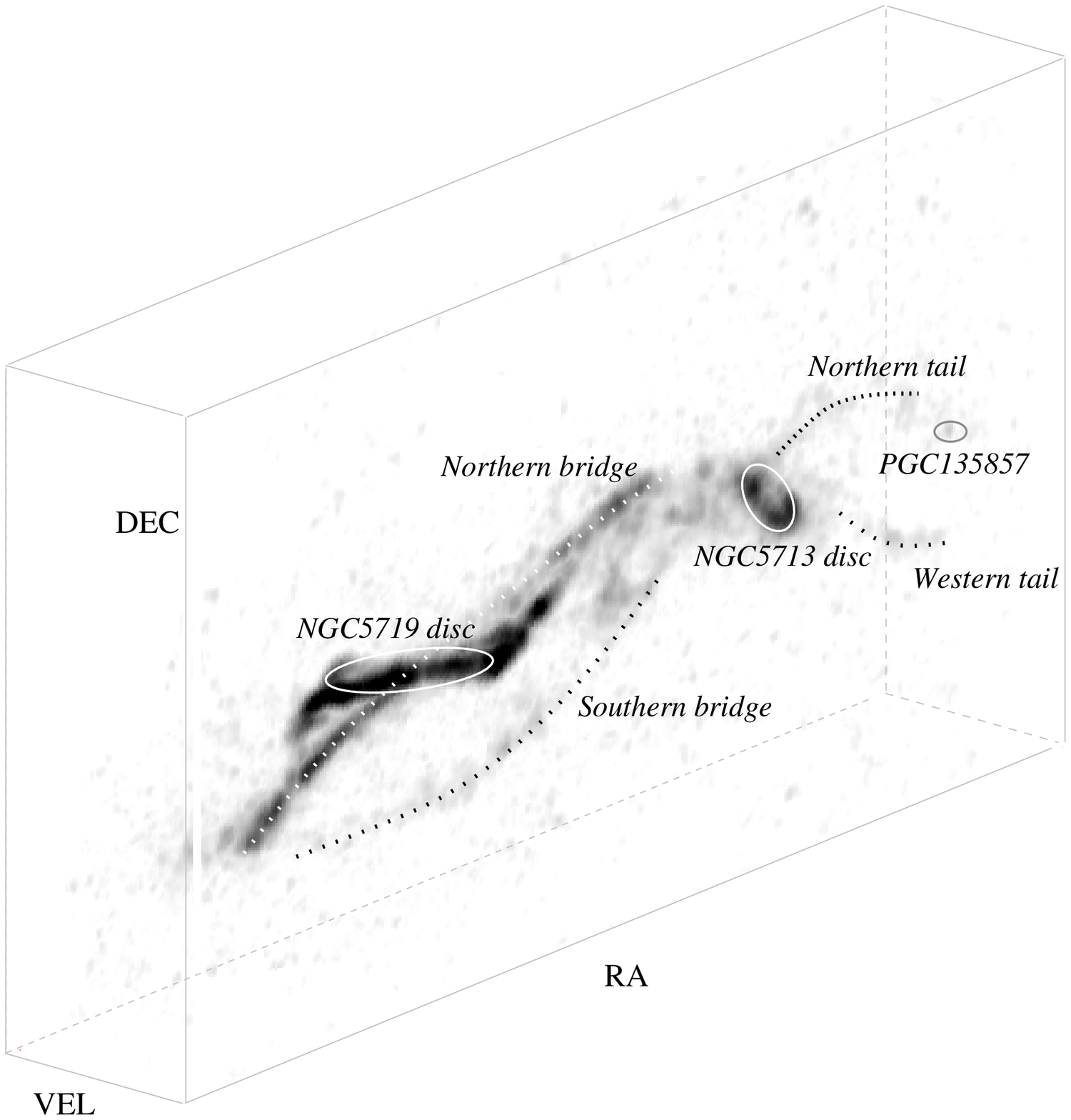}
\end{center}
\vskip 0.5cm
\caption{A three-dimensional representation of the \hi\ emission in the
 NGC\,5719/13 galaxy pair with a different viewing angle than
 Fig.\,\ref{fig:mom0velfi}. Here the cube is rotated by 20\degr\ around the
 declination-axis and by 50\degr\ around the velocity-axis to better visualise
 the different structures (marked in the figure).}
\label{fig:3d}
\end{figure*}

 % ===================== THE PGC COMPANION =====================
\begin{figure*}[t!]
\begin{center}
\leavevmode
\vskip -2.5cm
\hspace*{-0.15cm}\includegraphics[angle=0,width=20cm]{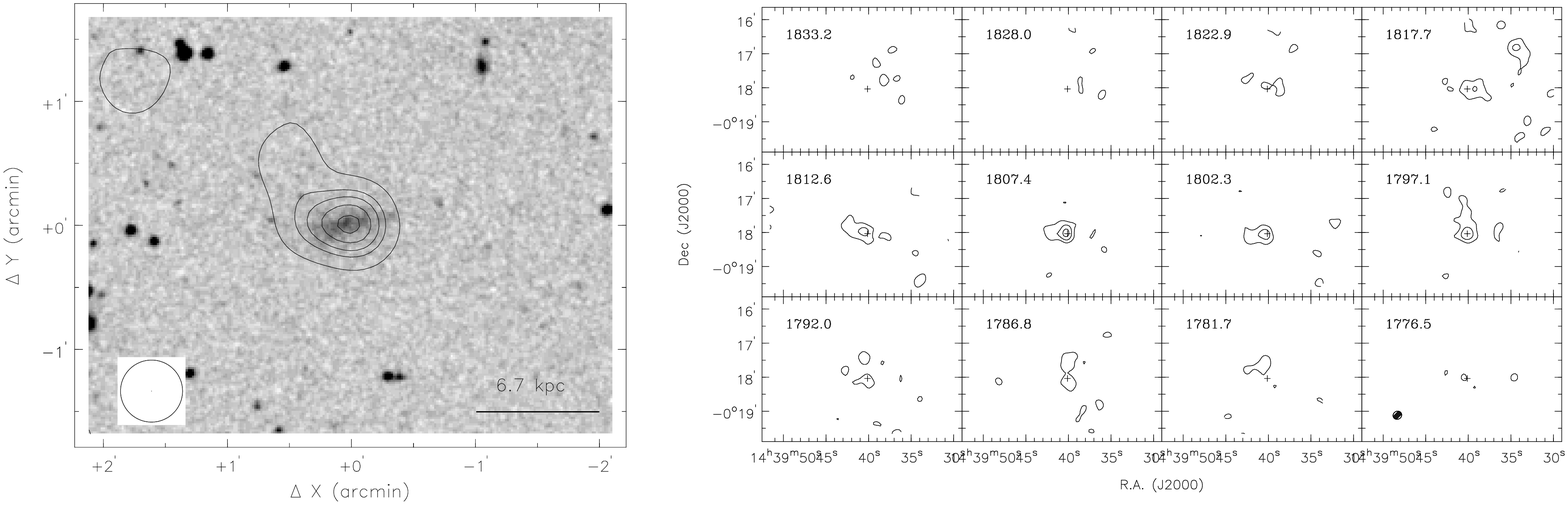}
\end{center}
\vskip -0.5cm
\caption{
({\it Left}) Contour map of the \hi\ column density distribution of the faint
  companion PGC\,135857 superimposed on an optical (DSS2) image. The lowest
  contour level is at $6\times1^{19}$ atoms~cm$^{-2}$ and the increment is
  $9\times10^{19}$ atoms~cm$^{-2}$. 
({\it Right})
Channel maps of the companion galaxy PGC\,135857. Contour levels are at $-1,
1$ ($\sim 2$~$\sigma$) to 3~\mjyb, in steps of 1~\mjyb. The spatial resolution
is $30\arcsec\times30\arcsec$. The cross represents the optical centre of
PGC\,135857.}
\label{fig:pgc}
\end{figure*}

 % ======== PVD ========
\addtocounter{figure}{+1}
\begin{figure*}[t!]
\begin{center}
\leavevmode
\vskip -0.5cm
\includegraphics[angle=-90,width=14cm]{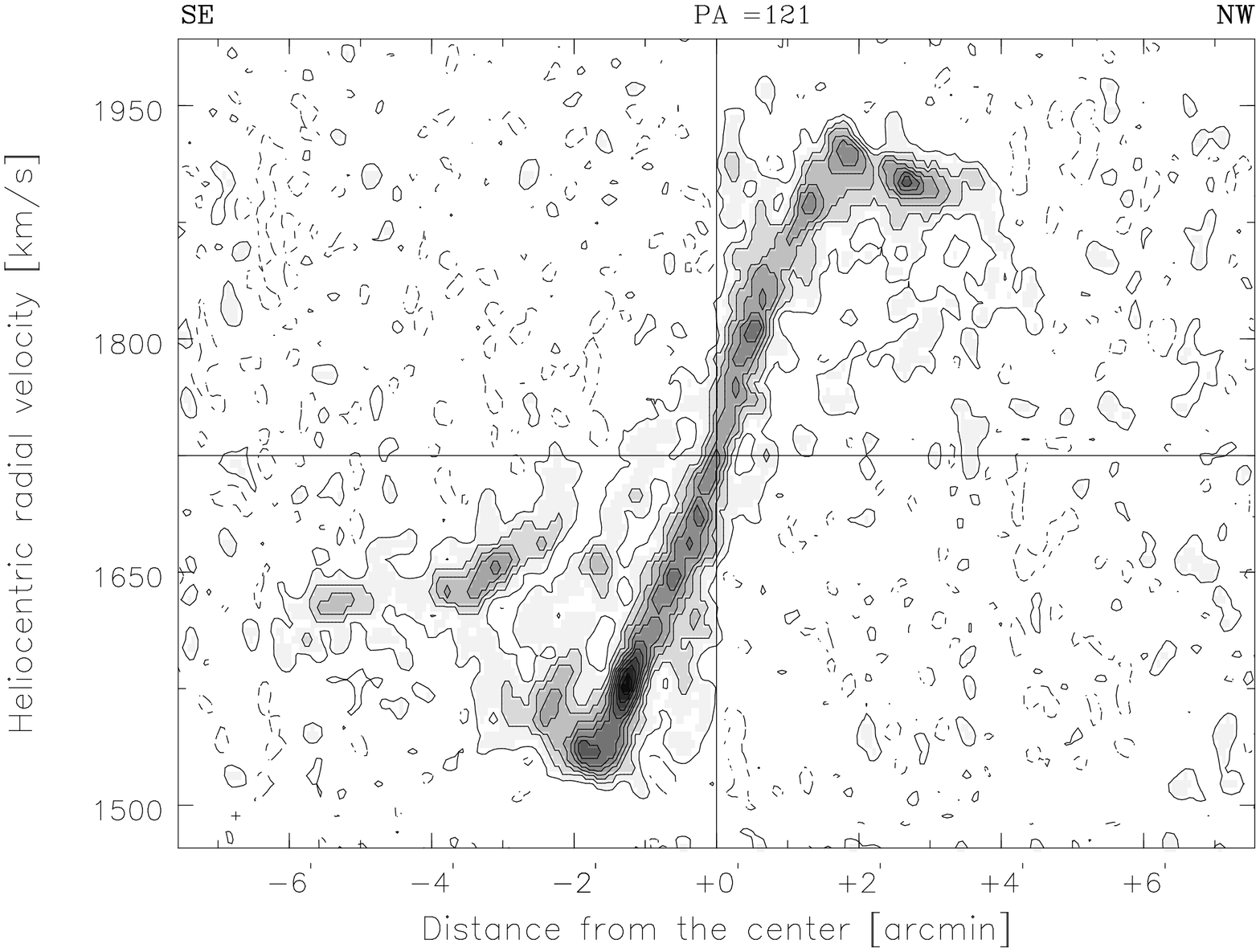}

\includegraphics[angle=-90,width=14cm]{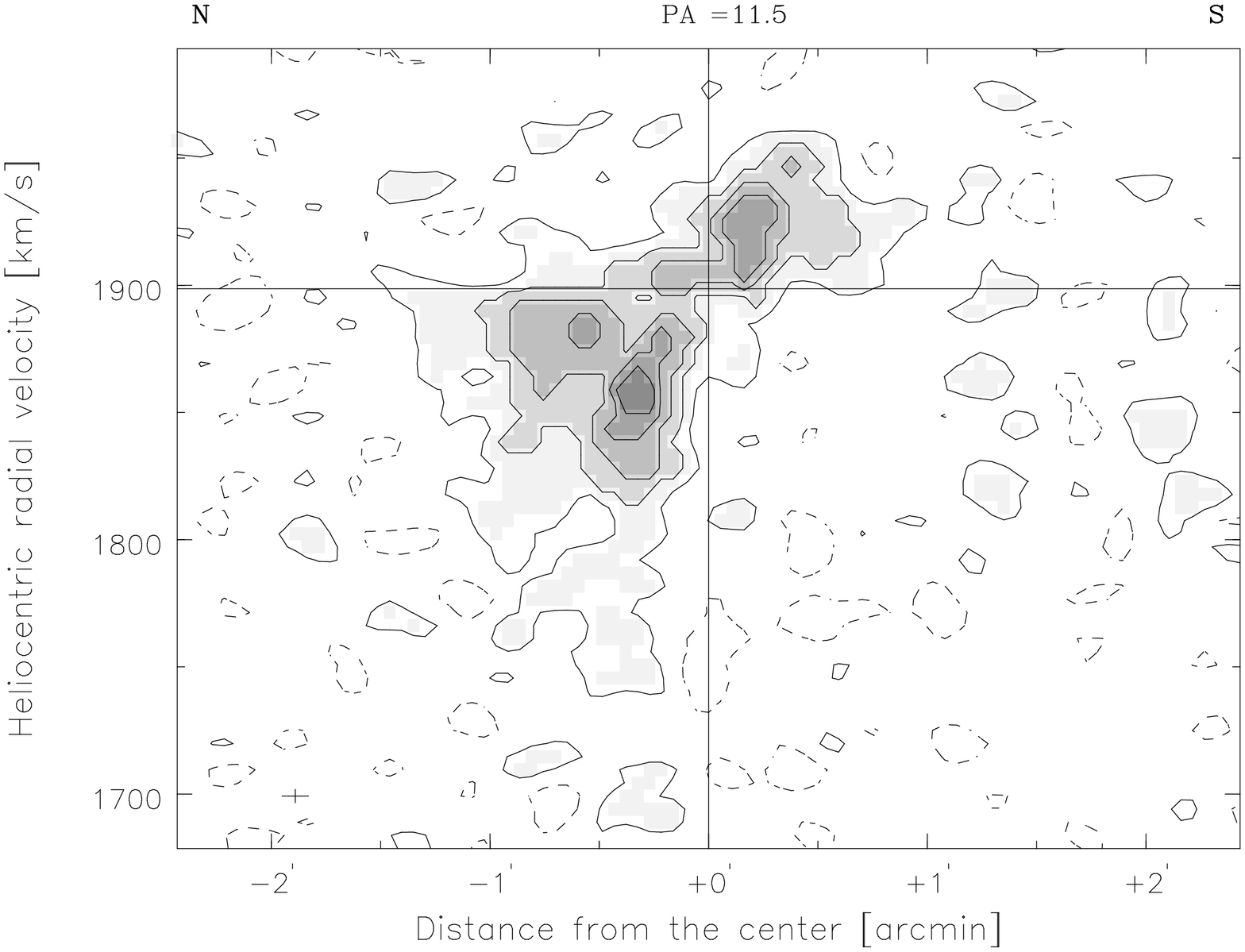}
\end{center}
\caption{Position-velocity diagrams along the \hi\ major axis of 
NGC\,5719 (${\rm P.A.} = 121\degr$, top) and NGC\,5713 (${\rm P.A.} = 11.5\degr$, bottom).
The horizontal lines indicate the systemic velocities of the galaxies. 
Contour levels are at -3, -1.5 (dashed), and from 1.5 to 27$\sigma$, in steps
of 1.5$\sigma$ (1$\sigma$~$\approx$ 0.5~mJy\,beam$^{-1}$). The angular and
velocity resolutions are 15\arcsec\ and 5.4~\kms, respectively (crosses at the
bottom left side).}
\label{fig:pvdhi}
\end{figure*}

\subsection{Optical spectroscopy}
 
We took optical long-slit spectra of NGC\,5719 at the European Southern
Observatory (ESO) in La Silla with the New Technology Telescope (NTT)
on July, 2002.
The ESO Multi-Mode Instrument (EMMI) was used in the Red Medium Dispersion
Spectroscopy (REMD) configuration. Grating No.~6 with 1200 grooves mm$^{-1}$
blazed at 5500~\AA\ was used in the first order in combination with a
1\farcs0~--~wide slit and the mosaiced MIT/LL CCDs No. 62 and 63 with
$2048\times4096$ pixels of $15\times15\;\mu$m each. The wavelength range
between 4800 and 5500~\AA\ was chosen to record simultaneously the \hb\ and
\oiii $\lambda$5007~\AA\ emission lines and the \mgi\ absorption triplet at
$\sim 5170$~\AA. The instrumental resolution, obtained by measuring the width
of sky lines on a science frame after the wavelength calibration, was
FWHM$=1.16$~\AA\ ($\sigma\approx 30$~\kms ) at 5100~\AA. The on-chip
$2\times2$ pixel binning provided a reciprocal dispersion of $0.4$~\AA\
pixel$^{-1}$ and an angular sampling of $0\farcs33$ pixel$^{-1}$. 

The slit was centred on the galaxy nucleus using the guiding TV
camera and aligned along the optical major axis ($\rm P.A.=96^\circ$). Two
spectra of 50 minutes each were obtained. 
A comparison lamp exposure was obtained after each object integration
to allow accurate wavelength calibration. Quartz lamp and twilight sky
flat-fields were used to remove pixel-to-pixel variations and
large-scale illumination patterns. Several G and K stars were observed
with the same set-up to serve as templates in measuring the stellar
kinematics.
The two spectra were bias subtracted, flat-field corrected, cleaned
from cosmic rays, wavelength calibrated, co-added and sky subtracted
using standard MIDAS\footnote{MIDAS is developed and maintained by the
European Southern Observatory} routines.

The stellar kinematics was measured from the galaxy absorption
features present in the wavelength range.
We used the Fourier Correlation Quotient method (FCQ, Bender 1990)
following the prescriptions of Bender et al. (1994). The spectra were
binned along the spatial direction to obtain a nearly constant
signal-to-noise ratio larger than 20 per resolution element. The
galaxy continuum was removed row-by-row from each spectrum by fitting
a fourth to sixth order polynomial. 
The template spectrum used was that of the K0.5 III star HR5196.
The line-of-sight stellar velocity ($v_\star$), and velocity dispersion
($\sigma_\star$) were determined by fitting a Gaussian to the line-of-sight
velocity distribution (LOSVD) at each radius. Velocities were corrected to
heliocentric. 
We derived errors on the stellar kinematics from photon statistics and
CCD read-out noise, calibrating them by Monte Carlo simulations as
done by Gerhard et al. (1998). These errors do not take into account
possible systematic effects due to template mismatch.

The ionised gas kinematics was measured by simultaneous Gaussian
fitting of the \hb\ and \oiii $\lambda$5007~\AA\ emission lines in the
resulting spectrum. The galaxy continuum was removed from the spectra, as was
done for measuring the stellar kinematics. We fitted in each row of the
continuum-subtracted spectrum a Gaussian to each emission line, assuming them
to have the same line-of-sight velocity ($v_g$) and velocity dispersion
($\sigma_g$). Velocities and velocity dispersions were corrected to
heliocentric and for instrumental velocity dispersion, respectively. Far from
the galaxy centre (for $|r|>20''$) we averaged adjacent spectral rows in order
to increase the signal-to-noise ratio of the emission lines. 
The uncertainties in the kinematic parameter determination were derived from
photon statistics and CCD read-out noise by means of Monte Carlo simulations.

% ======== HI + OPTICAL PVD ========
\begin{figure*}[t!]
\begin{center}
\leavevmode
\vskip -1cm
\includegraphics[angle=0,width=20.0cm]{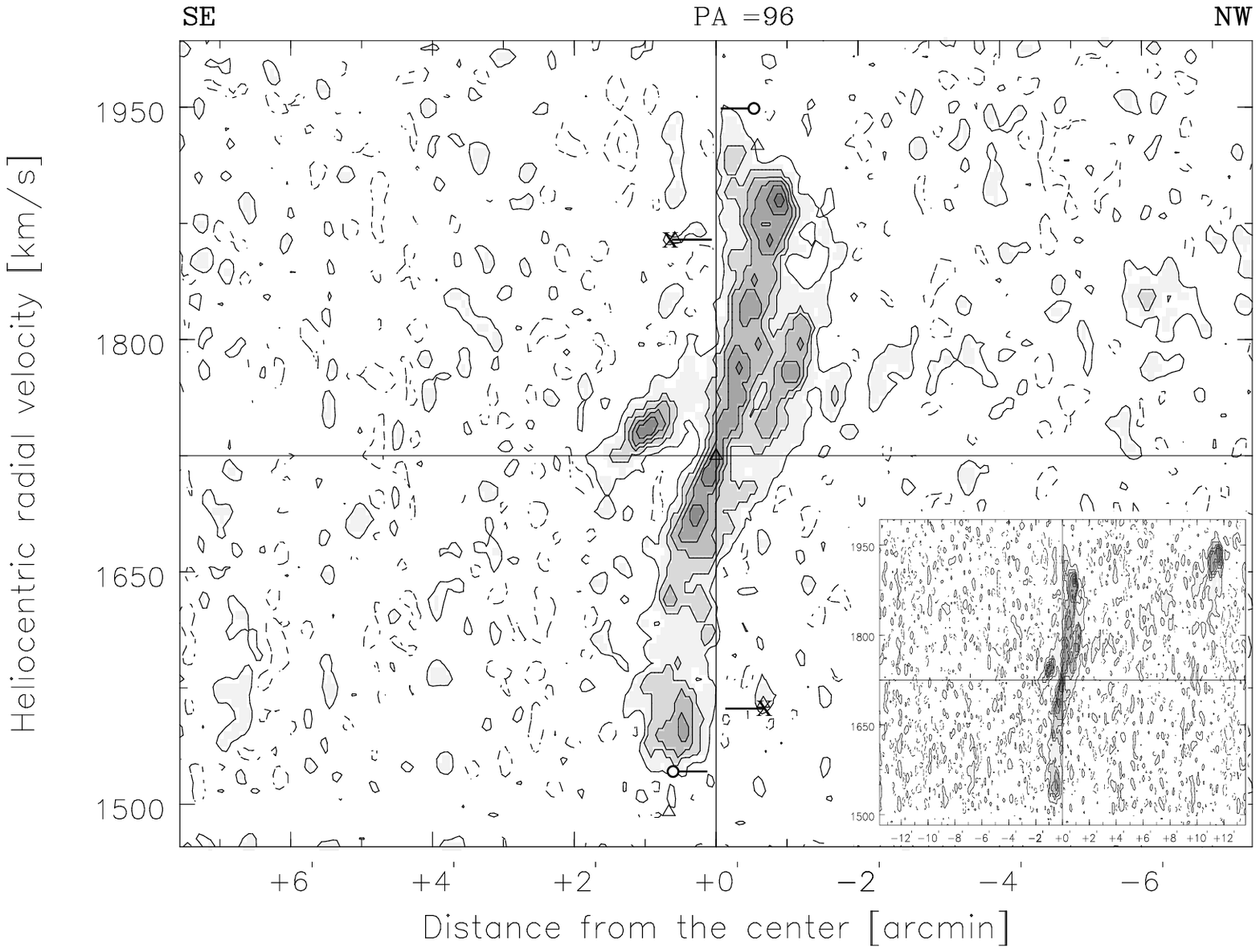}
\end{center}
\caption{ 
\hi\ position-velocity diagram obtained along the ${\rm P.A.} =96\degr$ optical
major axis of NGC\,5719, with overlaid the rotation curves of the main stellar
disc (crosses), the ionised gas (circles), and counter-rotating stars
(triangles). 
({\it Main panel}): 
The horizontal line indicates the systemic velocity of NGC\,5719.
The contour levels are at -3, -1.5 (dashed), and from 1.5 to 27$\sigma$ in
steps of 1.5$\sigma$ (1$\sigma$~$\approx$ 0.5~mJy\,beam$^{-1}$). The angular
and velocity resolutions are 15\arcsec\ and 5.4~\kms, respectively. 
({\it Inset panel:}) \hi\ position-velocity diagram (${\rm P.A.} = 96\degr$) with a
larger radial coverage to show the emission from NGC\,5713 at $+$11\arcmin\
from the centre.} 
\label{fig:pvdopt}
\end{figure*}

% ====================== OPTICAL IMAGES ======================
\begin{figure*}
\begin{center}
\leavevmode
\includegraphics[angle=0,width=18cm]{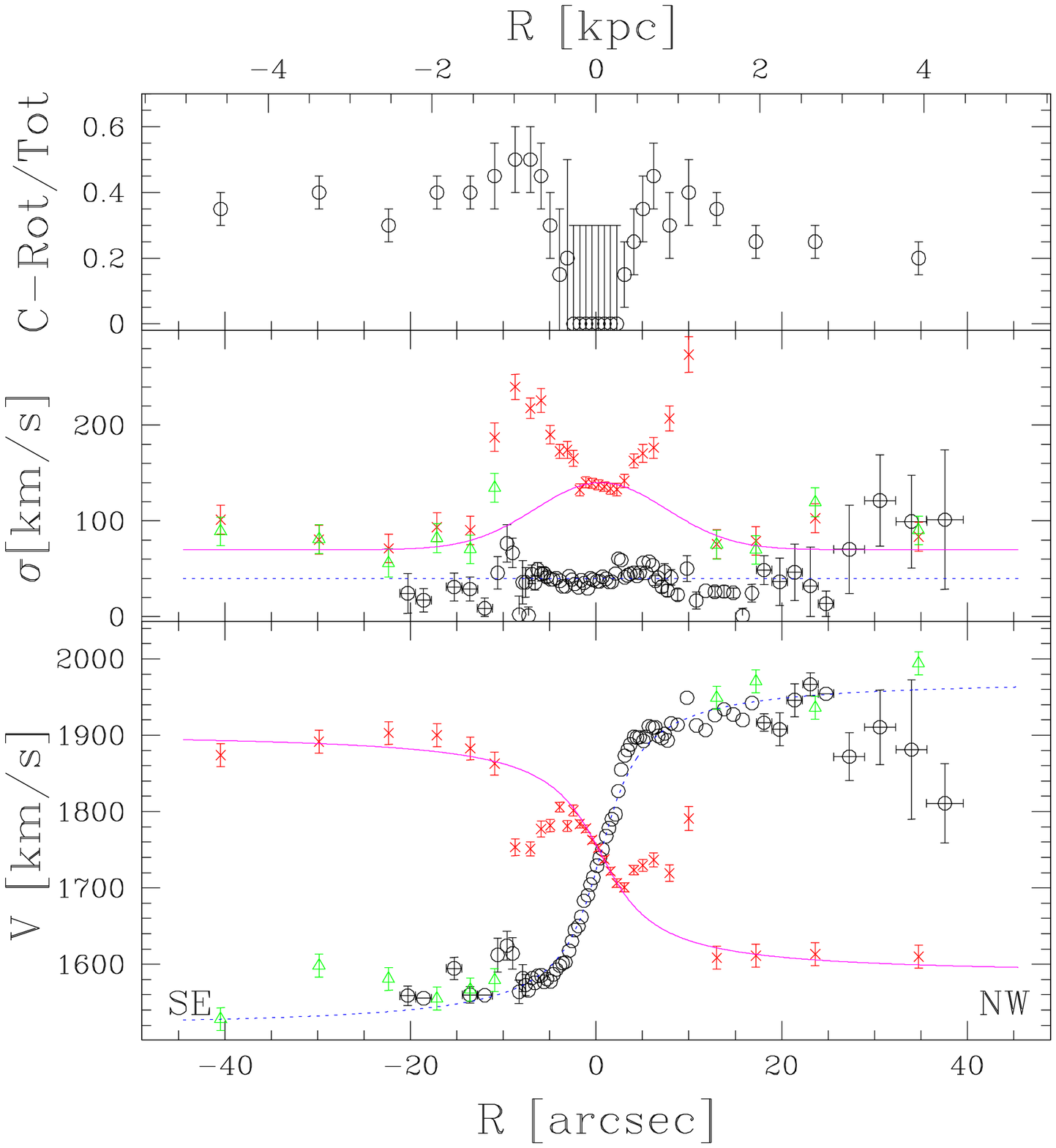}
\end{center}
\vskip -0.3cm
\caption{({\it Bottom panel}) 
Heliocentric radial velocities as function of position along the optical major
axis ($\rm P.A. =96^\circ$) of the ionised gas ({\it circles}), the main stellar
component ({\it red crosses}) and the counter-rotating stars ({\it green triangles}).
({\it Middle panel}) The velocity dispersion of the different components: co- 
and counter-rotating stellar populations, and the gaseous component (see details 
in the Bottom panel).
({\it Top panel}) 
The luminosity fraction of counter-rotating stars as function of radius in 
NGC\,5719; see the text for details. }
\label{fig:rcopt}
\end{figure*}

\section{Results}
\label{sec:results}

In this section we describe the distribution and kinematics of the 
stellar and the cold+ionised gaseous components of the NGC\,5719/5713 system, 
in particular the on-going merging event in the galaxy pair and the detection 
of two counter-rotating stellar discs in NGC\,5719.

\subsection{HI morphology and kinematics}
 \label{sec:hi_morphology}

The most striking features of the \hi\ morphology are due to the on-going
interaction between NGC\,5719 and NGC\,5713, which is mostly unnoticeable at
optical wavelengths (Fig.\,\ref{fig:mom0velfi}).

In NGC~5719 the \hi\ extends to over five times its optical size on both
sides, reaching a radius of 8\arcmin\ or 54~kpc at the assumed distance of
23.2\,Mpc, compared to the 180~kpc overall \hi\ extent of the pair. The major
axis of its \hi\ distribution follows that of the oblique dust lane (${\rm
  P.A.} =121^\circ$), and both are skewed by 25$^\circ$ to that of the optical
body (${\rm P.A.}=96^\circ$). The gas associated with its disc is regularly
distributed within the optical boundaries and bends away beyond $\sim
1\farcm1$, forming an integral-shaped warp. The NW side is receding
(Fig.\,\ref{fig:mom0velfi}).

In its NW outermost optical disc the \hi\ distribution peaks where a chain of
faint optical condensations is seen. In particular, the clumps number 1 and 2
in Neff et al. (2005) coincide with \hi\ condensations at radial velocities of
$\sim 1637$~\kms\ and $\sim 1529$~\kms, respectively.
 
Its \hi\ kinematics are globally quite chaotic, especially outside the 
optical body, as commonly found in on-going mergers.

Inspection of the 3D data cube (Fig.\,\ref{fig:3d}) shows that all \hi\
structures, which are seen in projection in the \hi\ maps, actually reside in
different planes. As a consequence of projection effects the velocity map
shows sharp discontinuities at, e.g., 1780~\kms\ and 1560~\kms\
(Fig.\,\ref{fig:mom0velfi}, lower panel).

From the nearly face-on \hi\ disc in NGC\,5713 a bridge points toward 
and connects to the NW tip of the \hi\ disc in NGC~5719, which is marked 
as {\it Northern bridge} in Fig.\,\ref{fig:3d}. It extends over 114~kpc on the
sky and has a large velocity gradient of 300~\kms.
South of the main \hi\ disc in NGC\,5719 lies a weaker and clumpy bridge,
marked as {\it Southern bridge} in Fig.\,\ref{fig:3d}, which starts from the
NW tip of the NGC\,5719 disc at $\sim$ 1650~\kms, loops around NGC\,5719 and
moves towards NGC\,5713 with receding velocities, assuming that the structure
is trailing. 

We also detected \hi\ emission from PGC\,135857, a small $B_t= 17.40$ Sm-type
satellite with v$_{sys}$=1802~\kms, $\sim 63.6$~kpc E of the centre of
NGC\,5713 (see Figure\,\ref{fig:pgc}). 
Although our observations are affected by primary-beam attenuation 
(FWHP$=30$\arcmin), we detected a fair amount of \hi\ in it ($M_{\hi}/L_B \sim
0.18$ M$_\odot/L_{\odot,B}$). 
In its close surroundings we detected two low column density \hi\ tails at
v$\sim$1800~--~1950~\kms: the {\it Northern} and {\it Western tails}, see
Fig.\,\ref{fig:3d}. The {\it Northern tail} is likely physically connected to
PGC\,135857, as shown by the channel maps Fig.\,4. %(Fig.\,\ref{fig:chans}). 

NGC\,5719 has a total \hi\ mass $M_{HI}$ of $7.2 \times 10^{9}$ M$_\odot$ and
a mean \hi\ surface density of 5.11\,\msun\,pc$^{-2}$ -- values which lie in
the lower quartile of the distributions determined for Sab/Sb galaxies,
although its $M_{\rm \hi}/L_B$ ratio of 0.98 M$_\odot/L_\odot,B$ is
significantly higher than the median value found by Roberts \& Haynes
(1994). NGC\,5713 has $M_{HI}$=$6.6 \times 10^{9}$ M$_\odot$, which is at the
lower limit of the average value for Sb/Sbc galaxies (Roberts \& Haynes
1994). 
For the tidal features we estimated an $M_{HI}$ of $2.3 \times 10^{8}$~M$_\odot$, 
i.e. about 15\% of that of the NGC\,5719/13 system, by visually
selecting these features in each channel map where they could be distinguished
from the main emission. This value is likely to be a lower limit due to
projection effects.

In Fig.\,\ref{fig:pvdhi} we present the \hi\ position-velocity diagrams (PVDs)
along the \hi\ major axes of NGC\,5719 (${\rm P.A.}=121$\degr) and NGC\,5713
(${\rm P.A.} =11.5\degr$). In both PVDs the kinematics are fairly chaotic,
indicating the signatures of several planes of \hi\ emission.
Figure\,\ref{fig:pvdopt} shows the PVD along the optical major axis of 
NGC\,5719 (${\rm P.A.} =96\degr$), as well as a zoomed-out version that shows the 74~kpc
long tail connecting NGC\,5719 and NGC\,5713. 
The maximum rotation velocities derived for the ionised gas, and the co- and
counter-rotating stars (see next Section) are shown for comparison. It is
clear that only a small fraction of the \hi\ in NGC\,5719 resides along the
optical major axis and that most of it lies along P.A.=121\degr. 

\subsection{Stellar and ionized gas kinematics}\label{sec:optical_kinematics}

\begin{figure}[t!]
\centering
\leavevmode
\vskip -0.3cm
\hspace*{-0.2cm}\includegraphics[angle=0,width=9.5cm]{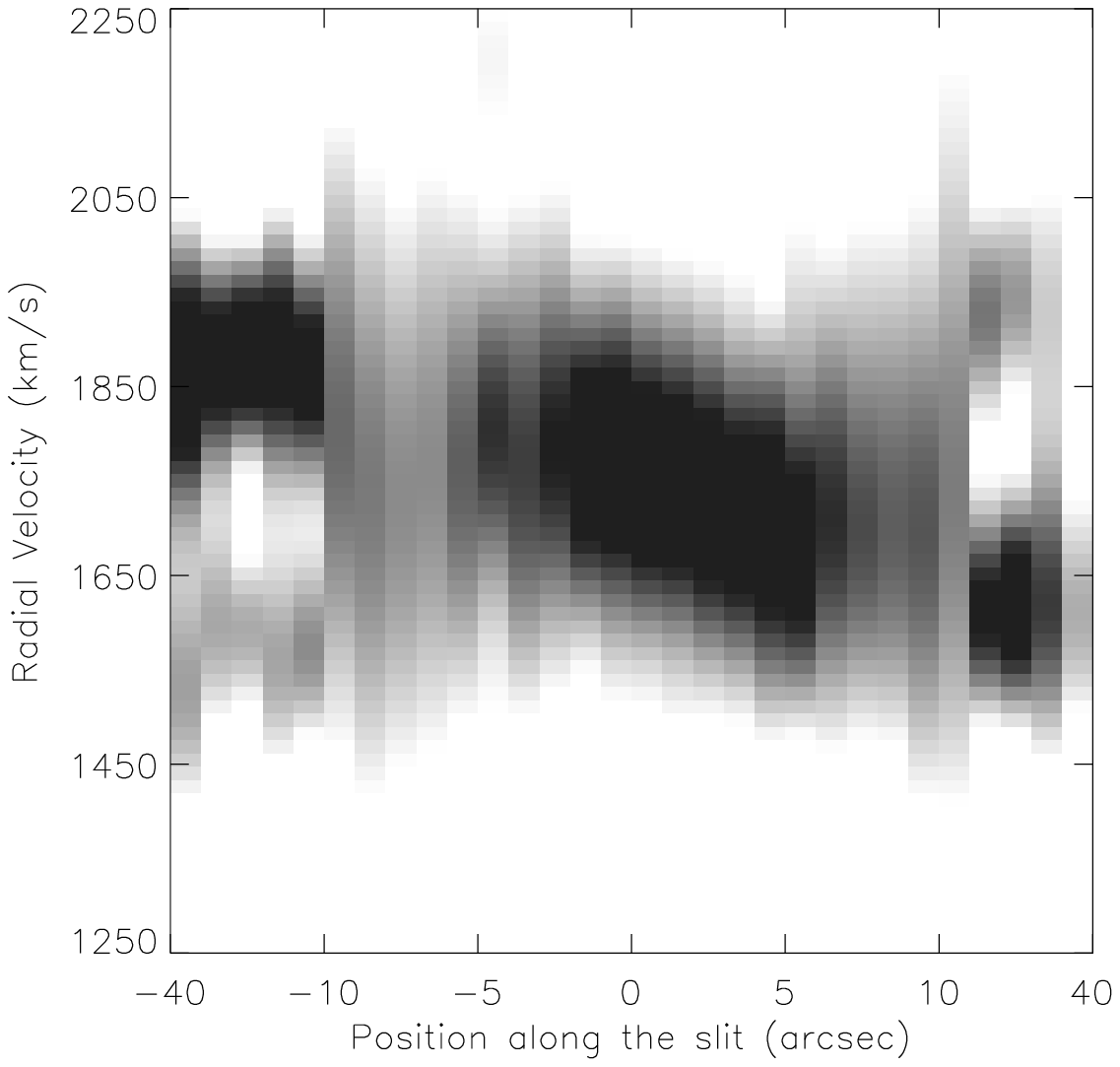}
\caption{Grey-scale plot of the stellar line-of-sight velocity 
distribution (LOSVD) in NGC\,5719, showing the projected stellar 
intensity along the optical ($\rm P.A. =96^\circ$) major axis, where the greyscale 
represents the stellar intensity, from zero (white) to maximum (black). 
The two counter-rotating stellar discs are clearly identifiable
at r$\ge$10\arcsec\ through the characteristic X-shaped pattern (see text).
For presentation purposes the spatial scale is not linear.}
\label{fig:losvd}
\end{figure}

The kinematics of the ionised gas and stars in NGC\,5719 are very complex. 
There are three distinct kinematical components. The rotation of the main
stellar component, which represents $\sim$80\% of the galaxy's light, is
receding towards the SE quadrant (Fig. \ref{fig:rcopt}) -- we have assumed
this as the reference sense of rotation of the galaxy. The ionised and \hi\
gas are rotating in the opposite sense. 
The presence of a second, counter-rotating stellar component is
revealed by the double-peaked stellar LOSVD (Fig.\ref{fig:losvd}).

At $|r|>10$\arcsec\ the main and counter-rotating components can be
disentangled due to their large difference in line-of-sight velocities
($\Delta v_\star \sim 350$~\kms) and small velocity dispersions ($\sigma_\star
< 100$~\kms ). The X-shaped pattern is the signature of counter-rotation (see
for comparison Fig.\,11.18 of Binney \& Merrifield 1998). At $|r| \sim
10$\arcsec\ $\Delta v_\star$ decreases and the two components blend with a
large LOSVD ($\sigma_\star \sim 240$~\kms ), which is likely due to the
combined $\Delta v_\star$ of the two unresolved components rather than to an
increase of their intrinsic velocity dispersion.
In this region we measured $v_\star\sim0$~\kms, indicating that the two
components have roughly the same intensity. 
At small radii the main component dominates the observed kinematics. 
The features observed in the radial profile of $\sigma_\star$ are also
indicative of the presence of a counter-rotating component.
In the region where the two components are unblended they both show 
low values of $\sigma_\star$, which we expect for a rotating stellar disc
($\sigma_\star \sim 80$~\kms ).
At $|r|<5$\arcsec\ $\sigma_\star$ is constant, suggesting that only one
component is present in this region.

We derived the relative intensity of the counter-rotating component as a
function of radius. We constructed a simple kinematical model by using the
spectrum of the stellar template to build the galaxy spectrum. We adopted for
the prograde and retrograde components the values of the line-of-sight
velocity and velocity dispersion we fitted to the stellar and ionised-gas
kinematics, respectively. 

For each radius a synthetic galaxy spectrum was obtained by co-adding the two
spectra we derived by convolving the stellar template spectra of the two
counter-rotating components. The relative intensities of the two components
were constrained by comparison of the $v_\star$ and $\sigma_{*}$ values
measured in the synthetic and observed spectra.
We found that for $|r|<3$\arcsec\ the data are consistent with the absence of
counter-rotating stars, although we can not exclude them contributing up to
$30\%$ of the total light. At $|r|\sim8''$ the counter-rotating component
reaches its maximum intensity ($\sim 50\%$ of the total light). At larger
radii its contribution decreases to $\sim 30\%$ on the SE side and $\sim
20\%$ on the NW side, respectively. 

Our observations are consistent with the measurements of the ionised-gas and
stellar kinematics by Rhee et al. (2004). Due to a lower spectral resolution
they measured no rotation in the very centre of NGC\,5719, but did not resolve
the counter-rotating component. Nevertheless, at large radii on the SE side
their data show the presence of stars rotating at $v_\star \la 100$~\kms\ in
an opposite direction than the ionised gas and corresponding to the most
luminous stellar component of NGC\,5719.

\section{Discussion}
\label{sec:discussion}

Both NGC\,5719 and NGC\,5713 are characterized by morphological and
kinematical peculiarities. 
NGC\,5719 has a skewed dust lane and shows faint condensations at its
optical edge, and NGC\,5713 appears very lopsided.
The on-going interaction between the galaxies is evident from the \hi\
distribution and kinematics, which show an intricate system extending over
more than 27\arcmin\ in different planes in the sky. This indicates that gas
was removed from NGC\,5713 by NGC\,5719 and/or stripped from their discs into
the vicinity, probably during different close passages. The kinematics of both
\hi\ discs are disturbed, particularly in the portions which are close to the
bridges that connect the two galaxies.
We speculate that a substantial fraction (at least $10\%$) of the entire \hi\
mass of the system that now resides in the tidal features around NGC\,5719,
may once have belonged to NGC\,5713, which is presently relatively gas-poor
for its morphological classification.
The presence of the small satellite galaxy PGC\,135857 seems to be related to
the interaction of the main galaxy pair, and it might have formed as a result
of a large local \hi\ reservoir being subjected to the physical conditions
that triggered star formation. 

NGC\,5719 hosts a counter-rotating component which contains both ionised and
neutral hydrogen, as well as $\sim 20\%$ of the disc stars. The
counter-rotating stellar disc has the same spatial extent as the main stellar
disc component.
Compared to the other known cases of spiral galaxies with counter-rotating
stellar discs, the counter-rotating stars in NGC\,5719 have the same radial
distribution as in NGC\,4138 (Jore et al. 1996) and NGC\,7217 (Merrifield \&
Kuijken 1994), and they rotate in the same sense as the gaseous component,
like in NGC\,4138 and NGC\,3593 (Bertola et al. 1996). All these
counter-rotating discs have the same fraction ($20$--$30\%$) of the total
luminosity of the host galaxy. On the other hand, NGC\,5719 is the only known
object belonging to an interacting system with stripping and/or gas accretion
gas that has stellar counter-rotating discs. This give us a unique opportunity
to shed light on the origin of a counter-rotating stellar disc, as the end
result of the star formation occurred in newly acquired material due to galaxy
interaction.

The ionised gas rotates in the same direction and in the same plane as the
\hi\ disc. They both counter-rotate with respect to the main stellar disc and
have nearly identical velocity amplitudes ($\Delta v \approx440$~\kms). The
velocity amplitude of the counter-rotating stars is close to that of the
neutral and ionised gas, while that of the main stellar disc is lower ($\Delta
v_\star \la 320$~\kms).

Given its kinematical properties, it is reasonable to invoke a galaxy
interaction as the origin of the stellar counter-rotation in the NGC\,5719
disc, rather than internal mechanisms.
The radial velocity amplitudes of the various components indicate that
the counter-rotating gaseous disc and the subsequent in-situ stellar
formation that occurred in it are attributable to the acquired \hi.

As in NGC\,4138 (Jore et al. 1996) the data suggest that the most of the stars
in the counter-rotating disc formed later from the ionised-gas disc, as commonly
found in normal galaxies.

Population synthesis analysis has been used (e.g., Carollo et
al. 1997) to determine the age difference between the co-rotating and
counter-rotating stellar components in kinematically-decoupled cores
in ellipticals. Recent results (Morelli et al. 2004; McDermid et
al. 2006) show that most of these cores, which are at most a few
hundred parsec in size, are younger than the rest of the host
galaxy. This favours the in-situ stellar formation scenario in galactic 
cores.
However, this technique has never been applied to large-scale ($>1$
kpc) counter-rotating discs, which are subject to different 
physical conditions and processes than galactic cores.
The spectroscopic resolution of our data does not allow this kind of
analysis. Moreover, the counter-rotating component of NGC\,5719 does not
dominate the light distribution at any radius, making measurements of its line
strength indices infeasible. We therefore adopted other diagnostics to
investigate the origin of stars in the counter-rotating disc.

Neff et al. (2005), combining the GALEX UV data with VLA \hi\ data obtained
using a configuration similar to ours (presented succinctly in Langston \&
Teuben 2001), already pointed out that there is a strict coincidence between
the location of the UV-bright regions and \hi\ column density peaks within and
far from the NGC\,5719 disc.
The stars dominating the UV light in the NGC\,5719 disc are very young
($300$--$400$~Myr), whereas other regions outside the main stellar disc appear
to be even younger ($2$--$200$ Myr), and there are many locations in the
northern and southern bridges where bright UV clumps were detected. 
The young age of the stellar population is consistent with the dynamical age
of the \hi\ bridges (e.g., Mihos \& Hernquist 1994, 1996), which probably
exceeds 500~Myr, the estimated epoch of the closest approach between the two
galaxies. This epoch was derived from their current projected separation
($\sim 80$~kpc) and assumed relative velocity ($\sim 200$~\kms).
Although we can not directly associate the young UV condensations with
the stars in the counter-rotating disc due to the 5-6\arcsec\ angular
resolution of GALEX (Morrissey et al. 2005) and projection effects, the
spatial \hi --UV association in the NGC\,5719 disc, bridges, and tails
indicate that the accreted gas in the counter-rotating disc is actually
forming a substantial number of stars. 
All these data support a scenario where the neutral hydrogen from the large
reservoir available in the surroundings was first accreted by NGC\,5719 onto a
retrograde orbit and subsequently formed in-situ into a counter-rotating
stellar disc.

We argue that PGC\,135857 was formed in the interaction process, as suggested
by the frequency with which such dwarf companions are observed in similar
cases like Stephan's Quintet (Mendes de Oliveira et al. 2004), the Antennae
(Hibbard et al. 2005), and other spectacular interactions (e.g., Neff et
al. 2005, Duc \& Mirabel 1998).
Even if we cannot confirm with the present data that it could have been 
pre-existing to the current interaction event, the Northern tail forms a 
physical connection between NGC\,5713 and PGC\,135857 and the 
Western tail has very similar velocities.

Concerning the evolution of the system, it is likely that the flow of
accreted matter will enable star formation to continue for time-scales
on the order of 1~Gyr after the tidal feature formation (Hibbard \&
Mihos 1995) and, as such, to have a strong impact on both the
evolution and the dynamics of NGC\,5719. The galaxy might eventually
build a secondary stellar disc as massive as the main one, and evolve
into an object similar to NGC\,4550 (Rubin et al. 1992; Rix et al. 1992).

The alternative scenario involving {\em only} the accretion of
already-formed stars is very unlikely due to the young age of the
bright UV stars observed in the \hi\ density peaks. However, we can
not reject the accretion on retrograde orbit of a low-mass satellite
that would {\em partially} contribute to the build up of the counter-rotating
stellar population.
The radial distribution of the counter-rotating component in NGC\,5719
can be described as a disc with a central hole. This is qualitatively
consistent with the results of the numerical simulations by Abadi et
al. (2003), where most of the thick counter-rotating disc was assembled
by stars from acquired and disrupted satellites. However, such a
scenario needs the {\em ad hoc} assumption of a past accretion of a
gas-rich dwarf satellite.
Two-dimensional optical spectroscopy that partially resolves the 
kinematics and maps the stellar age of the components should provide 
a definitive answer to this question.

\section{Conclusions}
\label{sec:conclusions}

The \hi\ distribution and kinematics show an on-going major merger 
between the edge-on Sab spiral NGC\,5719 and its Sbc spiral companion 
NGC\,5713. Tidal features in the form of two bridges that loop around
NGC\,5719 connecting to NGC\,5713, and two tails departing from NGC\,5713
towards the West were detected. The total extent on the sky of the \hi\
emission is about 27\arcmin, corresponding to $\sim 180$~kpc at the adopted
distance.
At the tip of the northern tail the low-mass satellite PGC\,135857 was
detected, which may represent a by-product of the interaction event.

The \hi\ column density distribution within and outside the NGC\,5719
disc is remarkably similar to the UV morphology, with a fairly precise
correspondence with the location of the clumps in a young stellar
population ($200$--$500$ Myr).

The neutral and ionised hydrogen in the disc of NGC\,5719 are counter-rotating
with respect to the main stellar disc. For the first time, a counter-rotating
stellar disc which contains nearly $\sim 20\%$ of the stars and which has the
same radial extension as the main stellar disc has been detected in an
interacting system.

The data support a scenario where neutral hydrogen was first 
accreted by NGC\,5719 onto a retrograde orbit from the large \hi\
reservoir available in the galaxy surroundings, and subsequently
formed in-situ into a counter-rotating stellar disc.

\section{Acknowledgments}
DV acknowledges supports by the European Commission through a Euro3D RTN on
Integral Field Spectroscopy (No. HPRN-CT-2002-00305) and a Marie Curie ERG
grant (No. MERG-CT-2005-021704).
This research has made use of the Lyon-Meudon Extragalactic Database
(HyperLeda), and the NASA/IPAC Extragalactic Database (NED) which is operated
by the Jet Propulsion Laboratory, California Institute of Technology, under
contract with the National Aeronautics and Space Administration.
The Very Large Array (VLA) is operated by the National Radio Astronomy
Observatory (NRAO). The NRAO is a facility of the National Science
Foundation operated under cooperative agreement by Associated
Universities, Inc.

% ====================== CHANNEL MAPS ======================
\addtocounter{figure}{-5} 
\begin{figure*}[h!]
\begin{center}
\leavevmode
\vskip -7cm
\hspace*{-1.5cm}\includegraphics[angle=0,width=23.0cm]{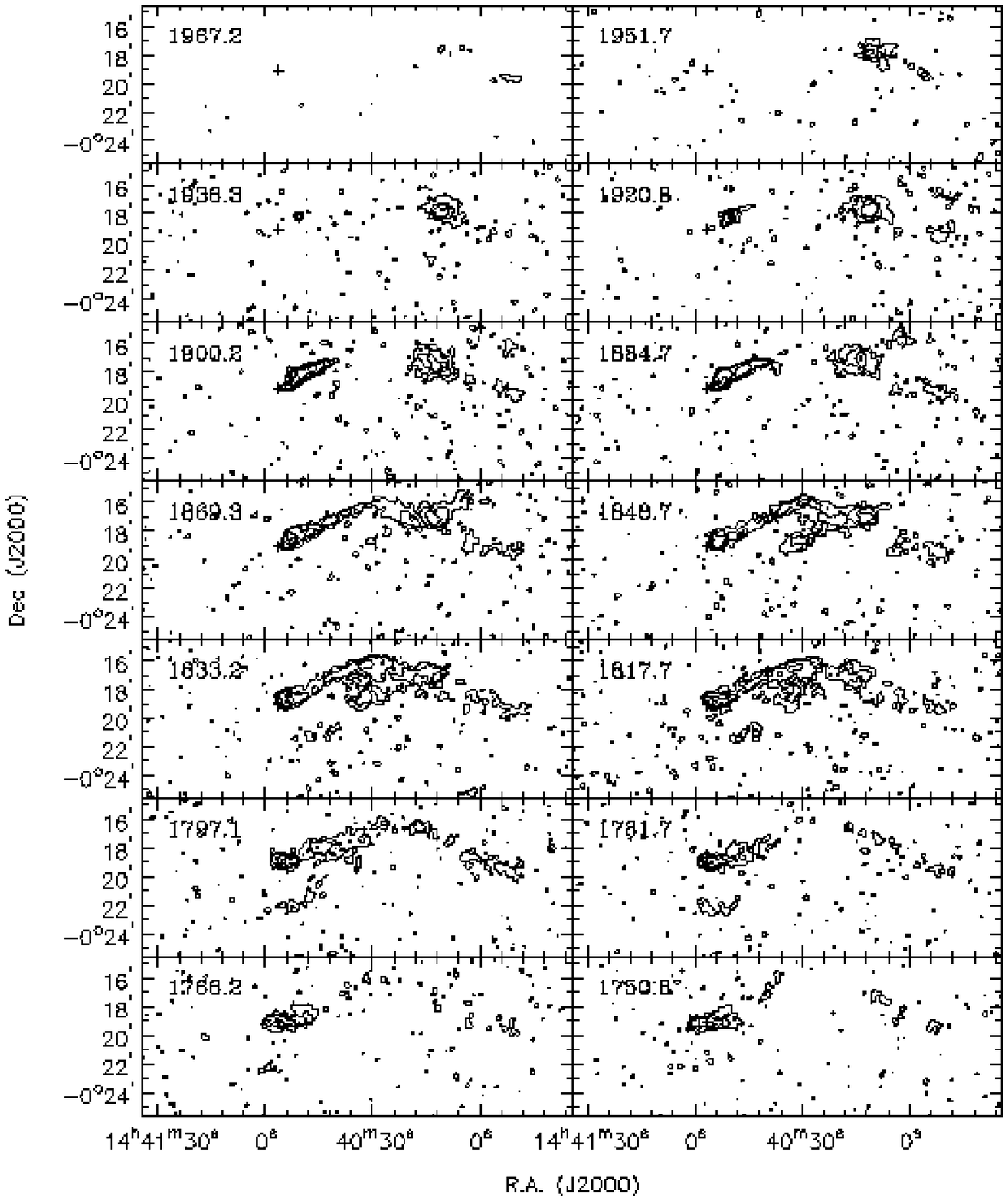}
\end{center}
\caption{Channel maps of the \hi\ emission in the entire NGC\,5719/13
  system. Contour levels are at -1.4, 1.4 ($\sim 2.5\sigma$), 2.8, 5.6, and
  11.2\mjb. The spatial resolution is $30\arcsec\times30\arcsec$. The cross
  represents the optical centre of NGC\,5719.}
\label{fig:chans}
\end{figure*}

\addtocounter{figure}{-1}
\begin{figure*}[h!]
\begin{center}
\leavevmode
\vskip -7cm
\hspace*{-1.5cm}\includegraphics[angle=0,width=23.0cm]{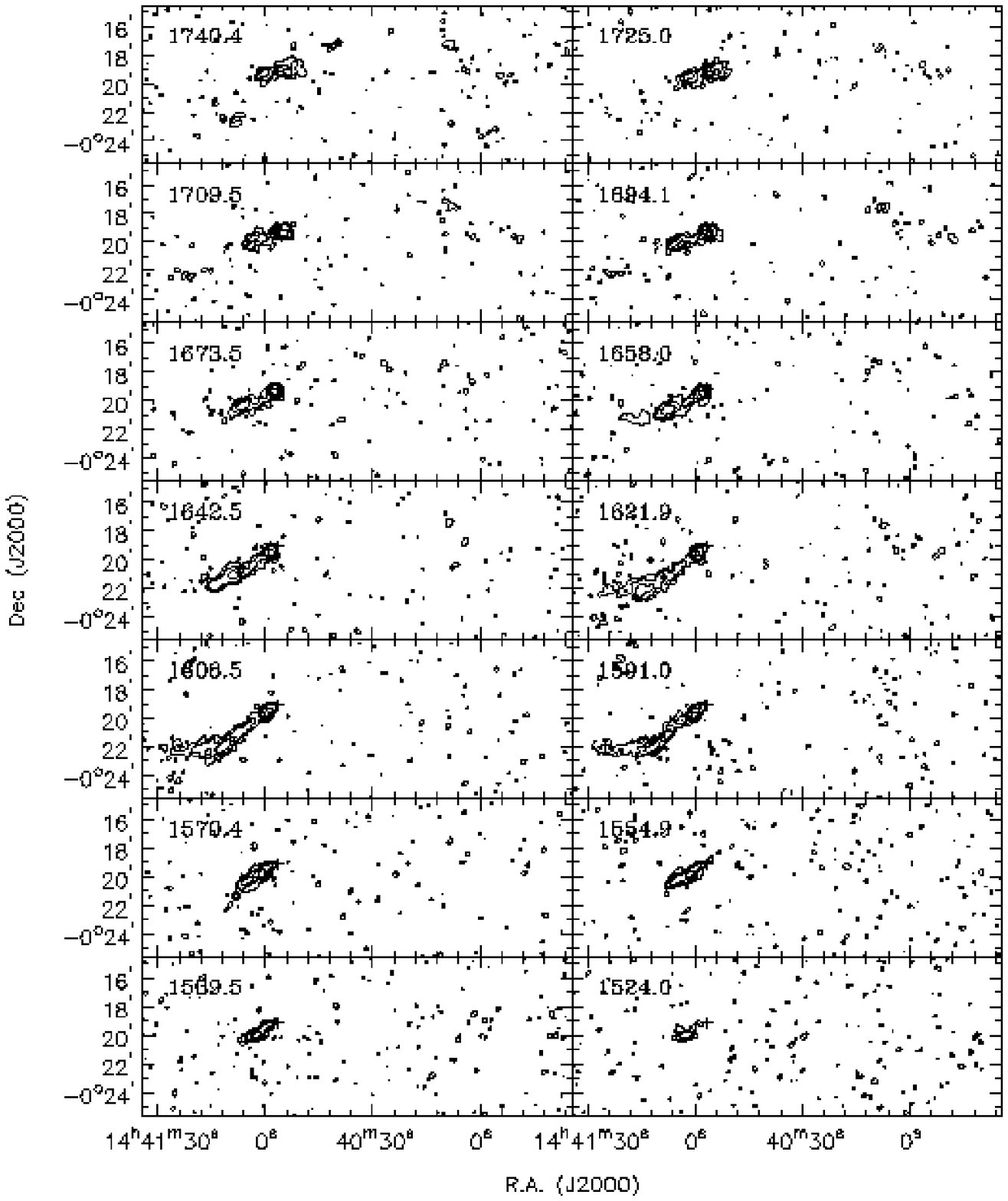}
\end{center}
{{\bf Fig.\,\ref{fig:chans}}. Continued.}
\end{figure*}

\end{document}